\documentclass[reprint,amsmath,amssymb,aps,prl,superscriptaddress,nofootinbib,floatfix]{revtex4-2}

\usepackage{graphicx}
\usepackage{dcolumn}
\usepackage{bm}
\usepackage{mathrsfs}
\usepackage{amsmath}
\usepackage{hyperref}
\usepackage{latexsym}
\usepackage{multirow}
\usepackage{booktabs}
\usepackage{microtype}
\usepackage{algorithm}
\usepackage{algpseudocode}
\usepackage{float}
\usepackage{tabularx}
\usepackage{array}
\usepackage{xcolor}

\newcolumntype{C}{>{\centering\arraybackslash}X}   
\newcommand{\etal}{et~al.}
\usepackage{tikz}
\usetikzlibrary{patterns,decorations.pathmorphing,shapes}
\hypersetup{colorlinks=true, linkcolor=blue, urlcolor=blue, citecolor=magenta}

\begin{document}

\title{Entropy-Driven Initiation and Cytoskeletal Viscoelasticity in Endocytosis: An Onsager Variational Framework}
\author{Jinjie Liu}
\affiliation{Institute of Theoretical Physics, Chinese Academy of Sciences, Beijing 100190, China}
\author{Zhongcan Ouyang}
\email{oy@itp.ac.cn}
\affiliation{Institute of Theoretical Physics, Chinese Academy of Sciences, Beijing 100190, China}
\author{Hao Wu}
\email{wuhao@ucas.ac.cn}
\affiliation{Zhejiang Key Laboratory of Soft Matter Biomedical Materials, Wenzhou Institute, 
University of Chinese Academy of Sciences, Wenzhou, Zhejiang 325000, China}

\date{\today}

\begin{abstract}
Receptor-mediated endocytosis requires a particle to approach the cell membrane to within a few nanometers before ligand--receptor binding can occur. Existing continuum models often start from an already established contact and do not explicitly describe how crowding particles on the extracellular side influence the distribution of the particle near the membrane. We examine entropic depletion forces as one possible nonspecific contribution to this initial approach. For ideal depletants, the Asakura--Oosawa excluded-volume construction gives an exact depletion potential for the planar geometry before contact. The potential and force vanish continuously at the onset of excluded-volume overlap. This interaction provides a possible contribution to membrane proximity before specific binding, while its extension to curved wrapping geometries requires additional approximation. Within a reduced continuum model, we combine depletion attraction, ligand--receptor binding, membrane deformation, and cytoskeletal viscoelastic dissipation. The viscoelastic contact is formulated through a hereditary integral and a standard linear solid. The kinetic model gives a conditional minimum ligand density for complete engulfment, a finite particle-size window, and a stiffness-dependent upper limit. When the stationary radius lies inside the domain of finite positive wrapping times, the estimated wrapping time has a minimum at a radius that decreases with increasing binding energy density. At fixed viscosity and other independent parameters, the same time approximation predicts slower wrapping as cell stiffness increases. The two positive roots defining the size window merge at a limiting parameter value, which characterizes closure of the admissible size interval. Depletion attraction is interpreted as one possible contribution to particle-membrane association, alongside electrostatic interactions, steric effects, and membrane fluctuations. The present analysis identifies how nonspecific attraction, specific adhesion, and mechanical resistance can contribute to different stages of membrane wrapping.
\end{abstract}

\maketitle

\section{Introduction}
\label{intro}

Endocytosis is a fundamental process by which eukaryotic cells internalize extracellular materials, including nutrients, viruses, and nanoparticles \cite{Doherty2009, zhang2015physical, wu2026unveiling}. Receptor-mediated endocytosis has been studied extensively, and receptor diffusion has traditionally been treated as a rate-limiting step \cite{Gao2005, Shi2008, mercer2010virus}. In this picture, free receptors diffuse to a binding site, form ligand--receptor complexes, and participate in membrane wrapping \cite{Marsh2006}. A complementary question concerns the initial proximity between the particle and the membrane. Ligand--receptor binding can occur only when the separation is within a few nanometers. Existing continuum models often assume that contact has already been established or that the particle is already close to the membrane. The influence of extracellular crowding particles on the distribution of the particle near the membrane is not explicitly resolved in these models.

{Recent experimental and theoretical studies indicate that endocytosis can require an initial affinity \cite{Verma2010}. Such affinity may arise from entropic depletion forces generated by smaller molecules, including globular proteins and polymer coils \cite{Nelson2004, Asakura1954, Asakura1958, Vrij1976}. Depletion effects are well established in colloid science. Larger particles attract when the excluded volumes of smaller particles overlap, which increases the accessible volume of the smaller \mbox{particles~\cite{Dinsmore1995, Imhof1995, Steiner1995, Ilett1995}}. Observations in vesicle systems \cite{Dinsmore1998} and in mammalian cells \cite{Irajizad2017} have been discussed in relation to depletion-driven membrane association. Reviews have emphasized entropy at the nano--bio interface \cite{Wan2024}. In crowded biological environments, crowding forces can drive macromolecular association without direct attraction \cite{Batra2009,Gupta2017,Dey2022,ellis2001crowding}. Entropic effects of membrane-anchored polymers can also modulate membrane mechanics and domain organization \cite{wu2013mechanical, wu2013effects}. Depletion forces are nevertheless relatively weak compared with specific ligand--receptor binding energies. Their primary role in endocytosis may be to reduce the effective barrier to initial contact rather than to drive wrapping by themselves.}

{A central geometric issue in applying depletion theory to endocytosis is the treatment of the depletion volume for finite-sized spheres near a planar membrane. The conventional spherical-cap approximation must be formulated in terms of the excluded volume of the small-particle centers. The correct volume for the planar geometry before contact is obtained from the overlap between the particle exclusion sphere of radius \(R+r\) and the membrane exclusion half-space \(z<r\). The resulting spherical cap is exact for the ideal hard-sphere Asakura--Oosawa model before contact. Its analytic continuation into the wrapping phase is a separate approximation and is not exact once the membrane deforms.}

The cell is not purely elastic. Beneath the membrane, the cytoskeleton, composed of actin, microtubules, and intermediate filaments in a viscous cytosol, confers viscoelasticity with creep and stress relaxation on timescales from seconds to minutes \cite{mofrad2009cytoskeletal, Rotsch1999}. These timescales can overlap with some endocytic events \cite{Ryan1996timing, Liou1997autophagic, Elkin2016endocytic}. Treating the cell as a linear elastic half-space \cite{Sun2006} or ignoring cytoskeletal deformation \cite{Gao2005} can therefore misrepresent the energy landscape. Viscoelastic resistance evolves over time and can act as a kinetic bottleneck. Experimental disruption of actin dynamics alters viral entry rates \cite{matarrese2005human}. The observed optimal uptake size around \(50\) nm \cite{Chithrani2006} has motivated models that include kinetic and non-equilibrium factors. Recent experiments have measured forces during cellular uptake of viruses and nanoparticles at the ventral side \cite{WiegandEtAl2020}, and viscoelastic properties of basal plasma membranes and cortices have been characterized \cite{Janshoff2021}.

Previous theoretical studies have incorporated cytoskeleton deformation to model engulfment kinetics after contact is established \cite{Zhang2021}, but they assume pre-existing contact and do not address initiation. Recent HIV modeling similarly assumes prior contact \cite{Kruse2023mathematical}. In this work, we examine entropic forces generated by nanoscale biomolecules as one possible contribution to the initiation step before specific binding. Once the particle is close, ligand--receptor binding sustains wrapping, and cytoskeletal viscoelasticity regulates kinetics through the Onsager variational principle \cite{Liu2026}. This principle minimizes the sum of the free-energy change rate and the Rayleigh dissipation function \cite{Onsager1931a, Onsager1931b, Doi2013}. It has been applied to polymer dynamics, colloids, and membranes \cite{Doi2013, Arroyo2009}. For engulfment, the generalized coordinate is the depth \(h(t)\), with driving force \(F(h)=-\partial E/\partial h\), leading to \(\zeta(h)\dot{h}=F(h)\). The elastic--viscoelastic correspondence principle \cite{Lee1960, Radok1957} extends Hertzian contact to viscoelastic media and incorporates time-dependent creep.

\textls[-25]{The model includes entropy-driven adhesion through the finite-size planar depletion geometry, membrane bending and tension through the Helfrich--Canham \mbox{Hamiltonian~\cite{Helfrich1973,Canham1970,Ouyang1989,wu2023,wu2025generalized}}, ligand--receptor binding, and cytoskeleton viscoelasticity through a standard linear \mbox{solid \cite{Johnson1985, Lee1960}}. The planar Asakura--Oosawa limit is recovered before contact. The model yields a conditional initiation concentration, a minimum ligand density, an engulfment size window, and an optimal particle radius that decreases with binding energy. Stiffer cells increase engulfment time and narrow the size window. The framework provides a reduced continuum description of cellular uptake with implications for virology, nanotechnology, and drug delivery.}

\section{Model and Methods}

\subsection{Initial Proximity and Entropic Driving Force}

\begin{figure}[H]
\includegraphics[width=0.5\textwidth]{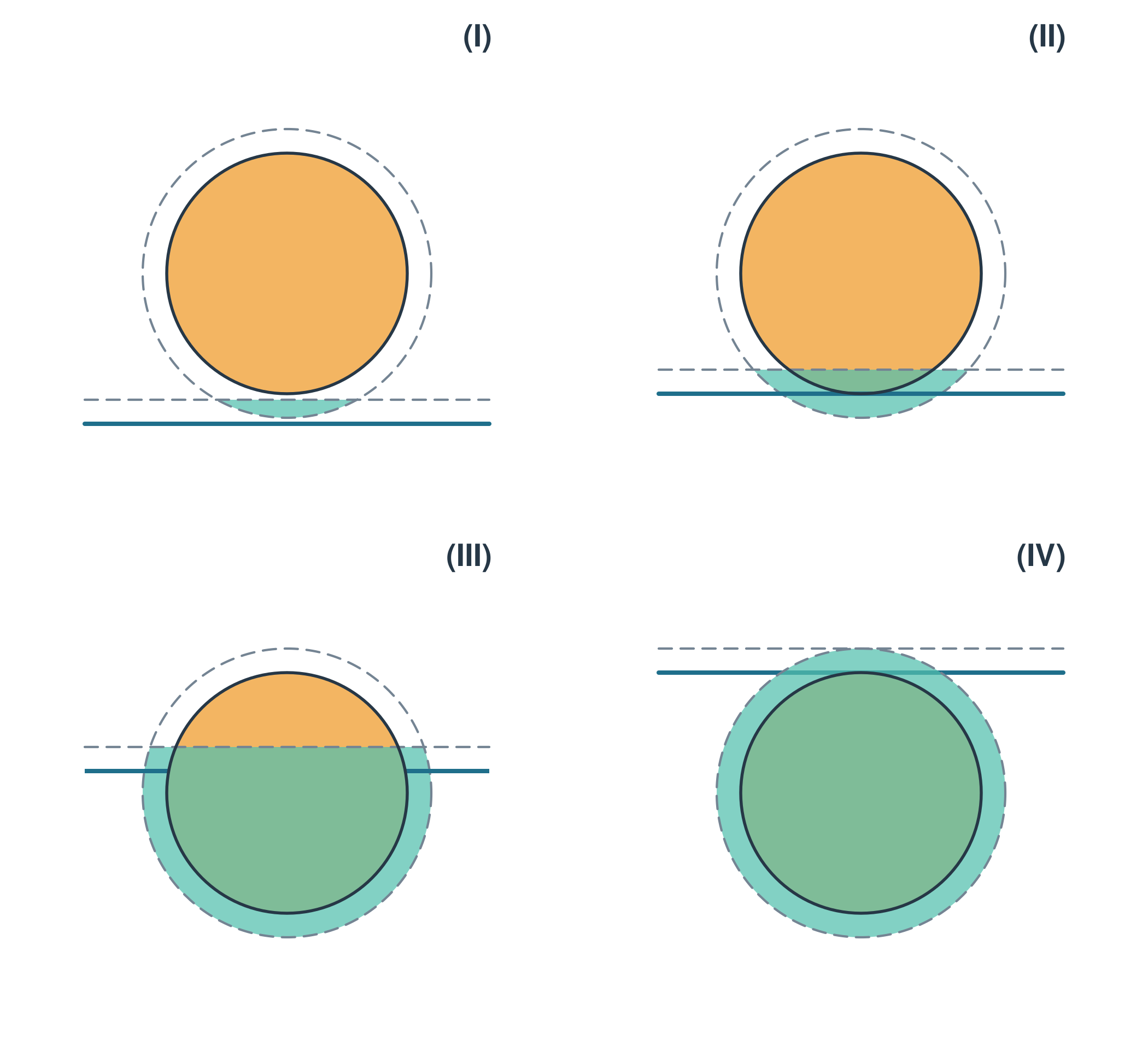}
\caption{
Entropy-induced engulfment model.
The yellow sphere represents the virus of radius \(R\).
The dashed spherical shell of thickness \(r\) around the virus is the depletion zone of the virus, i.e., the region excluded to the centers of crowding agents.
The region between the dashed and solid horizontal lines represents the depletion zone of the membrane.
The green area indicates the consumed (overlapped) portion of the depletion zones at the current engulfment stage; equivalently, it is the extra volume released to the external environment for the crowding agents to move freely.
This overlap increases the accessible volume for crowders and generates an effective attraction.
}
\label{fig:schematic}
\end{figure}

Consider a spherical viral particle of radius \(R\) suspended in a crowded environment containing smaller biomolecules of radius \(r\) at number density \(c\) (Figure \ref{fig:schematic}). In the absence of specific interaction, the virus and the cell membrane are separated by a distance determined by thermal motion. Ligand--receptor binding requires that the virus approach the membrane to within a distance where ligands and receptors can interact, typically a few nanometers. The present model examines one possible force that contributes to this approach.
\vspace{-6pt}

A depletion effect provides a possible contribution. Each large object, including the virus and the cell membrane, is surrounded by a depletion zone of thickness \(r\) into which the centers of smaller particles cannot enter. When the virus approaches the membrane, the depletion zones overlap. A region of excluded volume is eliminated, and the smaller particles explore a larger available volume. The entropy of the smaller particles increases, and the free energy of the system decreases. This generates an effective attractive force between the virus and the membrane.

{This entropic force can act without specific ligand--receptor binding. It may contribute to the initial driving force that brings the virus into proximity with the membrane, enabling subsequent specific interactions. Depletion forces alone are typically insufficient to overcome the membrane bending resistance for wrapping. Their role is primarily to facilitate the approach step. The crowding particles that produce this attraction are located on the extracellular side of the membrane. Intracellular crowding alone does not establish the same depletion mechanism at the extracellular membrane surface.}

\subsection{Planar Depletion Volume for the Ideal Asakura--Oosawa Model}

{We derive the depletion volume for the planar geometry using the ideal hard-sphere Asakura--Oosawa model. The key quantity is the overlap volume between the depletion zones surrounding the virus and the membrane. This construction is exact before contact for a planar membrane. Its extension to the wrapping phase is an approximation and is discussed separately.}

Let the membrane surface be at \(z=0\), occupying the half-space \(z<0\). The virus is a sphere of radius \(R\) centered at \((0,0,R+d)\), where \(d\ge0\) is the surface-to-surface separation. Small particles of radius \(r\) cannot penetrate either the virus or the membrane. The centers of crowders are excluded from two regions. The first region is a sphere of radius \(R+r\) centered at \((0,0,R+d)\), corresponding to the virus depletion zone. The second region is the half-space \(z<r\), corresponding to the membrane depletion zone after accounting for the crowder radius. The allowed region for crowder centers is \(z\ge r\).

{The overlap between the two excluded regions is a spherical cap of the sphere of radius \(R_{\rm eff}=R+r\), lying below the plane \(z=r\). The height of this spherical cap is
\begin{equation}
h_{\rm cap}=2r-d,
\end{equation}
{valid for \(0\le d\le 2r\). When \(d=2r\), the cap height is zero and overlap begins. When \(d=0\), the virus touches the membrane and the cap height is \(2r\).}

The volume of a spherical cap of height \(h\) in a sphere of radius \(R_{\rm eff}\) is
\begin{equation}
V_{\rm cap}(h)=\frac{\pi h^2}{3}(3R_{\rm eff}-h).
\end{equation}
Substituting \(h=2r-d\) and \(R_{\rm eff}=R+r\), we obtain
\begin{equation}
\begin{aligned}
V_{\rm dep}(d)&=\frac{\pi(2r-d)^2}{3}[3(R+r)-(2r-d)]\\
&=\frac{\pi}{3}(2r-d)^2(3R+r+d),
\label{eq:Vdep_d}
\end{aligned}
\end{equation}
{valid for \(0\le d\le 2r\). For \(d\ge 2r\), \(V_{\rm dep}=0\).}

{The osmotic pressure of the crowding agents is taken as the ideal van't Hoff relation}
\begin{equation}
P=ck_BT,
\label{eq:vant_Hoff}
\end{equation}
{where \(k_B\) is the Boltzmann constant and \(T\) is the absolute temperature. This approximation neglects non-ideal interactions, polydispersity, and spatial heterogeneity. Non-ideal corrections are discussed in Section \ref{sec5}.}

The entropic free-energy reduction for the planar geometry is
\begin{equation}
E_{\rm dep}^{\rm planar}(d)=-P V_{\rm dep}(d)=-ck_BT V_{\rm dep}(d).
\label{eq:Edep_planar}
\end{equation}

The depletion force along increasing \(d\) is
\begin{equation}
\begin{aligned}
F_d^{\rm planar}(d)&=-\frac{\partial E_{\rm dep}^{\rm planar}}{\partial d}\\
&=P\frac{\partial V_{\rm dep}}{\partial d}\\
&=-\pi P(2r-d)(2R+d).
\label{eq:Fdep_planar_d}
\end{aligned}
\end{equation}
{The negative sign indicates attraction toward decreasing \(d\). In the downward signed coordinate \(h=-d\), the planar approach branch is}
\begin{equation}
V_{\rm dep}^{\downarrow}(h)=\frac{\pi}{3}(2r+h)^2(3R+r-h),
\quad -2r\le h\le0,
\label{eq:Vdep_down}
\end{equation}
{and}
\begin{equation}
\begin{aligned}
F_h^{\rm planar}(h)=P\frac{\partial V_{\rm dep}^{\downarrow}}{\partial h}
=\pi P(2r+h)(2R-h)>0, \\
-2r\le h\le0.
\label{eq:Fdep_down}
\end{aligned}
\end{equation}
{This force is positive along increasing \(h\), which corresponds to decreasing separation \(d\).}

\subsection{Wrapping-Phase Depletion Adhesion}

\begin{table*}[t]
\caption{Parameters used in the model. References are provided for each parameter.}
\label{tab:parameters}
\begin{tabular*}{\textwidth}{@{\extracolsep{\fill}}l l l@{}}
\toprule
\textbf{Parameter} & \textbf{Value} & \textbf{References} \\
\midrule
Virus radius \(R\) & \(50\)--\(100\) nm & \cite{Chithrani2006,Burrell2016} \\
Crowder radius \(r\) & \(5\)--\(20\) nm & \cite{Batra2009,Gupta2017,Kuznetsova2014,Dey2022,ellis2001crowding} \\
Crowder concentration \(c\) & \(1.5\times10^{22}\) m\(^{-3}\) & \cite{Mitchison2019,Alfano2024} \\
Receptor-ligand complex size \(\delta\) & \(5\) nm & \cite{Gao2005} \\
Total adhesion energy density \(a\) & \(2.9\)--\(5.8\times10^{-4}\) J\,m\(^{-2}\) & \cite{Sun2006,chen2022quantification} \\
Boltzmann constant \(k_B\) & \(1.38\times10^{-23}\) J\,K\(^{-1}\) & Fundamental constant \\
Model temperature \(T\) & \(300\) K & Model temperature \\
Membrane bending modulus \(\kappa\) & \(8.18\times10^{-20}\) J & \cite{Li2012} \\
Membrane surface tension \(\gamma\) & \(2.07\times10^{-5}\) N\,m\(^{-1}\) & \cite{Sun2006,Li2024,Dai1998} \\
Water viscosity \(\mu\) & \(10^{-3}\) N\,s\,m\(^{-2}\) & Standard value \\
Cytoskeleton viscosity \(\eta_c\) & \(2696\) N\,s\,m\(^{-2}\) & \cite{Micoulet2005} \\
Virus Young's modulus \(E_v\) & \(10^8\) N\,m\(^{-2}\) & \cite{Sun2006,Li2012} \\
Cell Young's modulus \(E_c\) & \(2\times10^3\) N\,m\(^{-2}\) & \cite{Micoulet2005,Muenker2024} \\
Creep retardation time \(\tau=\eta_c/E_c\) & \(1.348\) s & \\
\bottomrule
\end{tabular*}
\end{table*}

{Once the membrane attaches to the particle, the local geometry is no longer a flat membrane. The planar depletion volume in Equation \eqref{eq:Vdep_down} can be continued analytically to \(h>0\), but this continuation does not represent the exact excluded-volume overlap for a curved membrane. A physically consistent reduced description is to use the planar Asakura--Oosawa potential for the approach phase and to use the local parallel-plate adhesion density for the attached membrane. For two parallel planes in contact with ideal depletants, the overlap of the two exclusion layers per unit area is \(2r\). The adhesion energy density is}
\begin{equation}
w_{\rm dep}=2Pr=2ck_BTr.
\label{eq:w_dep}
\end{equation}
{For a spherical-cap contact area \(A_c=2\pi Rh\), the depletion adhesion energy in the wrapping phase is}
\begin{equation}
\begin{aligned}
E_{\rm dep}^{\rm wrap}(h)=-w_{\rm dep}A_c=-4\pi R h ck_BTr,\\
 <h\le2R.
\label{eq:Edep_wrap}
\end{aligned}
\end{equation}
{The corresponding constant force along increasing \(h\) is}
\begin{equation}
F_{\rm dep}^{\rm wrap}=2\pi R w_{\rm dep}=4\pi R ck_BTr,
\qquad 0<h\le2R.
\label{eq:Fdep_wrap}
\end{equation}
{In the numerical estimates below, the total adhesion energy density \(a\) is treated as the sum of ligand--receptor and local depletion contributions when a single effective adhesion density is used. This avoids double counting of the depletion contribution in the wrapping~phase.}

\subsection{Approach Dynamics with Wall Hydrodynamics}

Before the particle contacts the membrane, depletion interaction can drive it toward the membrane. In the overdamped limit, the motion is governed by
\begin{equation}
\zeta_{\rm eff}(h)\dot{h}=F_{\rm dep}(h),
\label{eq:approach_dynamics}
\end{equation}
where \(\zeta_{\rm eff}(h)\) is the effective friction coefficient.

{For a sphere approaching a planar wall at separation \(d\), the Stokes friction coefficient is modified by wall effects \cite{Brenner1961}:}
\begin{equation}
\zeta_{\rm wall}(d)=6\pi\mu R\lambda(d).
\label{eq:wall_friction}
\end{equation}
{The far-field expansion is}
\begin{equation}
\lambda(d)=1+\frac{9R}{8d}+\mathcal{O}\left(\frac{R}{d}\right)^2,
\qquad \frac{R}{d}\ll1.
\label{eq:wall_lambda_far}
\end{equation}
{For the approach phase in the present model, \(d\sim r\sim10\)--\(20\) nm and \(R\sim50\) nm, giving \(d/R\sim0.2\)--\(0.4\). The condition \(R/d\ll1\) is not satisfied. The far-field expansion is therefore used only as an order-of-magnitude estimate. The full Brenner solution gives \(\lambda\approx4.611\) at \(d/R=0.3\), rather than the far-field estimate \(4.75\). Near contact, for a rigid no-slip wall, the lubrication limit is}
\begin{equation}
\zeta_{\rm wall}(d)\sim\frac{6\pi\mu R^2}{d},
\qquad d\to0^+.
\label{eq:lubrication}
\end{equation}
{This lubrication resistance diverges as \(d\to0\). If the deterministic equation is integrated to geometric contact \(d=0\), the approach time diverges unless a finite capture distance is~specified.}

{The time required for the particle to traverse the approach phase from \(h_0=-2r+\epsilon\) to \(h=-d_c\) with \(0<d_c<2r-\epsilon\) is}
\begin{equation}
t_{\rm approach}=\int_{-2r+\epsilon}^{-d_c}\frac{\zeta_{\rm eff}(h)}{F_{\rm dep}(h)}\,dh.
\label{eq:t_approach_general}
\end{equation}
{Using the bulk Stokes friction \(\zeta=6\pi\mu R\) and Equation \eqref{eq:Fdep_down}, the bulk result for \(d_c=0\) is}
\begin{equation}
t_{\rm approach}^{\rm bulk}
=\frac{3\mu R}{P(R+r)}
\ln\!\left[\frac{r(2R+2r-\epsilon)}{R\epsilon}\right].
\label{eq:t_approach_analytic}
\end{equation}
{Equation \eqref{eq:t_approach_analytic} is the bulk expression. For a finite capture distance \(d_c>0\), the corresponding bulk expression is}
\begin{equation}
t_{\rm approach}^{\rm bulk}(\epsilon,d_c)
=\frac{3\mu R}{P(R+r)}
\ln\!\left[\frac{(2r-d_c)(2R+2r-\epsilon)}{\epsilon(2R+d_c)}\right].
\label{eq:t_approach_bulk_dc}
\end{equation}
{With wall corrections, the time is}
\begin{equation}
t_{\rm approach}^{\rm wall}(\epsilon,d_c)
=\frac{6\mu R}{P}
\int_{d_c}^{2r-\epsilon}
\frac{\lambda(d)}{(2r-d)(2R+d)}\,dd.
\label{eq:t_approach_wall_integral}
\end{equation}
{A constant multiplication by five is not valid over the whole interval because \(\lambda(d)\) depends on \(d\). In the no-slip lubrication limit, the integral diverges as \(d_c\to0\). A finite capture distance is therefore required for a finite wall-corrected approach time.}

{Using the parameter values in Table \ref{tab:parameters} with \(\epsilon=0.01\) nm and the bulk expression, we~obtain}
\begin{equation}
t_{\rm approach}^{\rm bulk}\approx 3.1\times10^{-4}\ {\rm s}.
\label{eq:t_approach_numeric}
\end{equation}
{The wall-corrected value depends on the capture distance. With a representative finite capture distance, it is of the order of \(10^{-3}\) s. The exact geometric contact time is not finite in the no-slip lubrication model.}

{The approach time depends logarithmically on the cutoff parameter \(\epsilon\). For \(\epsilon\) ranging from \(0.01\) nm to \(1\) nm, the logarithmic factor changes by a factor of approximately \(2.4\). The approach time is therefore relatively insensitive to the precise value of the cutoff, although the absolute value changes by more than a factor of two.}

{At the onset of depletion-zone overlap, \(h=-2r\), the depletion force is zero. The particle enters the depletion zone by thermal diffusion. Since the depletion force increases continuously from zero, there is no energy barrier in the depletion potential itself. Once inside, the particle experiences an increasing attractive force. The deterministic equation does not describe the entry process from outside the depletion zone. A stochastic description is required for the probability and residence time.}

{The calculated approach time is not the residence time in the binding-competent configuration. The approach time describes arrival at a capture distance. Stable binding also requires a sufficient residence time and a finite bond-formation rate. The comparison with molecular dynamics binding times \cite{hu2013binding} provides only an order-of-magnitude reference.}

{The depletion energy at contact for the planar model is}
\begin{equation}
|E_{\rm dep}(0)|=\frac{4\pi c k_B T r^2(3R+r)}{3}.
\end{equation}
{For representative parameters \(c=1.5\times10^{22}\) m\(^{-3}\), \(R=50\) nm, and \(r=15\) nm,}
\begin{equation}
|E_{\rm dep}(0)|\approx9.7\times10^{-21}\ {\rm J}\approx2.3\,k_BT.
\end{equation}
{This value is comparable to the thermal energy. It provides a modest stabilization of the particle near the membrane. Other contributions, including the glycocalyx, electrostatic interactions, and membrane fluctuations, may also contribute to the initiation of contact.}

{For hard-sphere crowders that cannot overlap, the volume fraction is}
\begin{equation}
\phi=\frac{4\pi}{3}cr^3.
\end{equation}
{For the representative value \(c=1.5\times10^{22}\) m\(^{-3}\) and \(r=15\) nm, \(\phi\approx0.21\). The ideal van't Hoff pressure is then a significant approximation. The Carnahan--Starling compressibility factor at \(\phi\approx0.21\) is approximately \(2.55\). Non-ideal pressure corrections can be of order one. If the crowders are treated as ideal noninteracting depletants, the ideal expressions remain a reference model. Replacing \(P\) by a non-ideal pressure alone does not give an exact non-ideal depletion potential.}

\subsection{Ligand--Receptor Binding Energy}

The virus-cell contact area during wrapping is \(A=2\pi Rh\). Let \(\rho_L\) denote the ligand surface density and \(e_{\rm RL}\) the energy per ligand--receptor bond. If all ligands in the contact region form bonds and the bond energy is counted as a negative contribution, the binding energy is
\begin{equation}
E_{\rm bind}(h)=-2\pi R h a,
\qquad a=\rho_L e_{\rm RL}.
\label{eq:Ebind}
\end{equation}
{The symbol \(\rho_L\) is used here to avoid conflict with the friction coefficient. The effective adhesion density \(a\) may include ligand--receptor and local depletion contributions when a single effective density is used. Recent studies have examined ligand mobility and configurational entropy in nanoparticle uptake \cite{ZhangLiWang2021PRE, ChenXueLiEtAl2024}, and steric interactions between mobile ligands can facilitate complete wrapping in passive endocytosis \cite{DiMicheleJanaMognetti2018}. These effects are not explicitly resolved in the continuum binding energy density. They influence the effective value of \(a\).}

\subsection{Membrane Deformation Energy}

The membrane deformation energy in the spherical-cap approximation is
\begin{equation}
E_{\rm mem}(h)=\frac{4\pi\kappa h}{R}+\gamma\pi h^2,
\label{eq:Emem}
\end{equation}
where \(\kappa\) is the bending rigidity and \(\gamma\) is the membrane surface tension. The first term represents the bending energy for wrapping a fraction of the particle. The second term represents the work against surface tension. {This expression is a spherical-cap approximation. It does not include the deformation energy of the membrane outside the contact region. Gauss curvature is assumed to cancel between compared states or to be absorbed into contact-line and constant contributions. Membrane tension also plays a dynamic role in the flat-to-curved transition during clathrin-mediated endocytosis \cite{BucherEtAl2018}, and adhesion energy controls lipid binding-mediated endocytosis \cite{GrozaEtAl2024}.}

\subsection{Cytoskeletal Viscoelasticity and Onsager Variational Formulation}

\subsubsection{Creep Compliance and Hereditary Integral}

The cytoskeleton exhibits viscoelastic behavior. For a standard linear solid, the creep compliance for the two-body contact is
\begin{equation}
\Phi(t)=\frac{1}{E_v}+\frac{1}{E_c}\left(1-e^{-t/\tau}\right),
\qquad \tau=\frac{\eta_c}{E_c},
\label{eq:Phi}
\end{equation}
where \(E_v\) is the virus Young's modulus, \(E_c\) is the cell Young's modulus, and for precision we choose the mean value over five cycle measurements from Table 2 in Ref. \cite{Micoulet2005} as the cytoskeletal viscosity \(\eta_c\). {The time \(\tau=\eta_c/E_c\) is a creep retardation time. It is not the stress relaxation time. For the same scalar network, the stress relaxation time under fixed total strain is \(\tau_{\rm rel}=\eta_c/(E_v+E_c)\). The first term in Equation \eqref{eq:Phi} represents the instantaneous elastic response of the virus. The second term represents the viscoelastic response of the cell. This combined compliance follows from the series connection of the virus and cell compliances~\cite{Johnson1985}. Experimental measurements of basal plasma membrane viscoelasticity~\cite{Janshoff2021} motivate the use of a standard linear solid as a model assumption.}

{For a viscoelastic half-space indented by a rigid sphere with a time-dependent force \(F(t)\), the hereditary integral is}
\begin{equation}
h_H^{3/2}(t)=\frac{9}{16\sqrt R}\int_{0^-}^{t}\Phi(t-s)\,dF(s),
\label{eq:hereditary}
\end{equation}
{when both materials are approximated as incompressible with Poisson ratio \(1/2\). For general Poisson ratios, the numerical prefactor is replaced by the appropriate reduced modulus combination. For a step-like force \(F(t)=F_0\Theta(t)\),}
\begin{equation}
h_H^{3/2}(t)=\frac{9F_0}{16\sqrt R}\Phi(t).
\label{eq:step_response}
\end{equation}
{This step-response relation is the quasi-static Hertzian contact result for the adopted incompressible approximation. The driving force in the present model varies with \(h\), so the hereditary integral should generally be used. The quasi-static approximation is used below as a reduced model and its limitations are stated.}

\subsubsection{Elastic Storage and Viscous Dissipation}

{The cytoskeleton stores elastic energy and also dissipates energy. It cannot be represented only by a local friction coefficient. For a rigid sphere and a Kelvin--Voigt cortex in the small-indentation Hertz limit, the contact force is}
\begin{equation}
F_c=\frac{16\sqrt R}{9}\left[E_c h_H^{3/2}+\eta_c\frac{d}{dt}h_H^{3/2}\right].
\label{eq:Fc_KV}
\end{equation}
{Since}
\begin{equation}
\frac{d}{dt}h_H^{3/2}=\frac{3}{2}\sqrt{h_H}\dot h_H,
\end{equation}
{Equation \eqref{eq:Fc_KV} becomes}
\begin{equation}
F_c=\frac{16}{9}E_c\sqrt R h_H^{3/2}
+\frac{8}{3}\eta_c\sqrt{Rh_H}\dot h_H.
\label{eq:Fc_KV_split}
\end{equation}
{The elastic force is}
\begin{equation}
F_{el}(h_H)=\frac{16}{9}E_c\sqrt R h_H^{3/2}.
\end{equation}
{The viscous friction coefficient is}
\begin{equation}
\zeta_{KV}(h_H)=\frac{8}{3}\eta_c\sqrt{Rh_H}.
\label{eq:zeta_KV}
\end{equation}
{The dimensions are}
\begin{equation}
[\eta_c\sqrt{Rh_H}]={\rm Pa\,s\,m}={\rm N\,s\,m^{-1}},
\end{equation}
{so \(\zeta_{KV}\) is a friction coefficient.}

{The elastic storage energy is}
\begin{equation}
U_{el}(h_H)=\int_0^{h_H}F_{el}(u)\,du
=\frac{32}{45}E_c\sqrt R h_H^{5/2}.
\label{eq:Uel}
\end{equation}

\subsubsection{Onsager Variational Equation}

{The generalized coordinate is the contact indentation \(h_H(t)\). The interaction free energy is}
\begin{equation}
E_0(h_H)=E_{\rm dep}(h_H)+E_{\rm bind}(h_H)+E_{\rm mem}(h_H).
\label{eq:E0}
\end{equation}
The generalized force conjugate to \(h_H\) follows from the interaction free energy in Equation~\eqref{eq:E0}. Differentiating the three contributions gives
\begin{equation}
\begin{aligned}
F(h_H)&=-\frac{dE_0}{dh_H}\\
&=F_{\rm dep}(h_H)+2\pi R a-\frac{4\pi\kappa}{R}-2\gamma\pi h_H.
\label{eq:F_h_general}
\end{aligned}
\end{equation}
Here \(F_{\rm dep}(h_H)=-dE_{\rm dep}/dh_H\) is the depletion contribution. For the approach phase \(-2r\le h_H\le0\), \(F_{\rm dep}(h_H)\) is given by Equation~\eqref{eq:Fdep_down}. For the wrapping phase \(0<h_H\le2R\), \(F_{\rm dep}(h_H)\) is given by Equation~\eqref{eq:Fdep_wrap} when the local depletion adhesion is treated separately. If \(a\) is instead defined as the total effective adhesion density that already includes the local depletion contribution, then the wrapping-phase depletion term is absorbed into \(a\) and should not be added separately; in that convention, \(F_{\rm dep}\) is retained only for the approach phase. The symbol \(h\) in the original image corresponds to \(h_H\) here.
{The total free energy including elastic storage is}
\begin{equation}
E_{\rm tot}(h_H)=E_0(h_H)+U_{el}(h_H).
\label{eq:Etot}
\end{equation}
{The Rayleigh dissipation function is}
\begin{equation}
\mathcal{R}(\dot h_H,h_H)=\frac{1}{2}\left[\zeta_f(h_H)+\frac{8}{3}\eta_c\sqrt{Rh_H}\right]\dot h_H^2,
\label{eq:Rayleigh}
\end{equation}
{where \(\zeta_f\) represents fluid or other non-cytoskeletal friction. The Onsager variational principle gives}
\begin{equation}
\frac{\partial E_{\rm tot}}{\partial h_H}
+\frac{\partial\mathcal{R}}{\partial\dot h_H}=0.
\label{eq:Onsager_var}
\end{equation}
{Therefore,}
\begin{equation}
\left[\zeta_f(h_H)+\frac{8}{3}\eta_c\sqrt{Rh_H}\right]\dot h_H
=-\frac{dE_0}{dh_H}-\frac{16}{9}E_c\sqrt R h_H^{3/2}.
\label{eq:Onsager_motion_correct}
\end{equation}
{The energy balance is}
\begin{equation}
\frac{dE_{\rm tot}}{dt}
=-\left[\zeta_f+\frac{8}{3}\eta_c\sqrt{Rh_H}\right]\dot h_H^2\le0.
\label{eq:energy_balance}
\end{equation}
{At equilibrium, \(\dot h_H=0\), and the balance is}
\begin{equation}
-\frac{dE_0}{dh_H}=\frac{16}{9}E_c\sqrt R h_H^{3/2}.
\label{eq:equilibrium}
\end{equation}
{The cell modulus \(E_c\) therefore affects the final indentation. This effect is absent when the cytoskeleton is placed entirely into a friction term.}

{If the finite virus modulus is retained, the single-exponential memory can be represented by an auxiliary variable \(Q(t)\) with dimensions of force. For the combined compliance Equation \eqref{eq:Phi}, one has}
\begin{equation}
\tau\dot Q+Q=F_c(t),
\qquad Q(0)=0,
\label{eq:Q_memory}
\end{equation}
\begin{equation}
h_H^{3/2}(t)=\frac{9}{16\sqrt R}\left[\frac{F_c(t)}{E_v}+\frac{Q(t)}{E_c}\right].
\label{eq:h_from_Q}
\end{equation}
{The contact force \(F_c\) is determined by the force balance with the driving force and any additional fluid friction. This memory formulation is exactly equivalent to the hereditary integral for the single-exponential kernel.}

\subsubsection{Wrapping-Time Expression}

\begin{table*}[t]
\caption{Stationary particle radius from the reduced wrapping-time expression as a function of adhesion energy density.}
\label{tab:opt_size}
\begin{tabular*}{\textwidth}{@{\extracolsep{\fill}}c c@{}}
\toprule
\textbf{Adhesion Energy Density} $\bm{a}$ \textbf{(J/m\(^2\))} & \textbf{Stationary Radius} $\bm{R_{\rm opt}}$ \textbf{(nm)} \\
\midrule
\(1.8\times10^{-4}\) & 59.5 \\
\(2.0\times10^{-4}\) & 55.6 \\
\(2.5\times10^{-4}\) & 48.5 \\
\(3.0\times10^{-4}\) & 43.6 \\
\(4.0\times10^{-4}\) & 37.0 \\
\(5.0\times10^{-4}\) & 32.7 \\
\(6.0\times10^{-4}\) & 29.6 \\
\bottomrule
\end{tabular*}
\end{table*}

{In the reduced quasi-static approximation, the contact relation for a step-like force is used with the instantaneous value of the driving force. For complete engulfment, the geometric indentation is written as \(h_H=2R\). The argument of the logarithm in the wrapping time becomes}
\begin{equation}
g(R)=1+\frac{E_c}{E_v}
-\frac{32\sqrt2 E_c R^2}{9F(2R)}.
\label{eq:g_R}
\end{equation}
{The wrapping time is}
\begin{equation}
t_c=-\tau\ln g(R).
\label{eq:t_c}
\end{equation}
{A positive and finite wrapping time requires}
\begin{equation}
0<g(R)<1.
\label{eq:g_domain}
\end{equation}
{This condition replaces the earlier one-sided inequality.}

{The driving force at complete engulfment is}
\begin{equation}
F(2R)=2\pi R a-\frac{4\pi\kappa}{R}-4\pi\gamma R.
\label{eq:F_2R}
\end{equation}
{Here \(a\) is the total effective adhesion energy density used in the reduced model. When the local depletion adhesion is treated separately, \(a\) is replaced by \(a+w_{\rm dep}\).}

{The upper bound \(g(R)<1\) gives}
\begin{equation}
F(2R)<\frac{32\sqrt2}{9}E_v R^2.
\label{eq:upper_bound}
\end{equation}
{The lower bound \(g(R)>0\) gives}
\begin{equation}
F(2R)>\frac{32\sqrt2}{9}E_\infty R^2,
\qquad
E_\infty=\left(\frac{1}{E_c}+\frac{1}{E_v}\right)^{-1}.
\label{eq:lower_bound}
\end{equation}
{Both bounds must be satisfied.}

\section{Results and Analysis}

\subsection{Conditions for Complete Engulfment}

{Substituting Equation \eqref{eq:F_2R} into the lower bound Equation \eqref{eq:lower_bound} gives the minimum adhesion energy density}
\begin{equation}
a_{\min}
=\frac{16\sqrt2}{9\pi}E_\infty R
+\frac{2\kappa}{R^2}
+2\gamma.
\label{eq:a_min}
\end{equation}
{In terms of ligand density,}
\begin{equation}
\rho_{L,\min}
=\frac{1}{e_{\rm RL}}
\left[
\frac{16\sqrt2}{9\pi}E_\infty R
+\frac{2\kappa}{R^2}
+2\gamma
\right].
\label{eq:rhoL_min}
\end{equation}
{When the long-time elastic resistance of the cytoskeleton is neglected, \(E_\infty\to0\), and}
\begin{equation}
a_{\min}^{\rm no\,elastic\,cyto}
=\frac{2\kappa}{R^2}+2\gamma.
\label{eq:a_min_no_cyto}
\end{equation}
{This limit coincides with the thermodynamic membrane criterion of Yuan \etal \cite{Yuan2010}. It does not imply that cytoskeletal viscous deformation disappears.}

For a typical virus radius \(R\sim50\) nm, using the parameters in Table \ref{tab:parameters}, the estimate is
\begin{equation}
a_{\min}\approx1.87\times10^{-4}\ {\rm J\,m^{-2}}.
\label{eq:a_min_numeric}
\end{equation}

\subsection{Size-Dependent Engulfment}

{The particle radius must lie within a finite range for complete engulfment. Setting the lower bound condition to equality and substituting \(F(2R)\) yields}
\begin{equation}
2\pi R a-\frac{4\pi\kappa}{R}-4\pi\gamma R
=\frac{32\sqrt2}{9}E_\infty R^2.
\label{eq:cubic_start}
\end{equation}
{Multiplying by \(R\) and rearranging gives}
\begin{equation}
\frac{32\sqrt2}{9}E_\infty R^3
+2\pi(2\gamma-a)R^2
+4\pi\kappa=0.
\label{eq:cubic}
\end{equation}
{The positive real roots \(R_{\min}\) and \(R_{\max}\) define the engulfment window}
\begin{equation}
R_{\min}<R<R_{\max}.
\label{eq:R_range}
\end{equation}
{The upper bound Equation \eqref{eq:upper_bound} must also be checked. In the parameter range used here, the upper bound does not further restrict the window.}

\subsection{Optimal Particle Size}

{The wrapping time exhibits a minimum at an optimal particle size when the optimum lies inside the domain of finite positive wrapping times. From Equation \eqref{eq:t_c},}
\begin{equation}
t_c=-\tau\ln g(R).
\end{equation}
{At fixed \(\tau\), the condition \(dt_c/dR=0\) gives \(dg/dR=0\). With}
\begin{equation}
g(R)=1+\frac{E_c}{E_v}
-\frac{32\sqrt2 E_c R^2}{9F(2R)},
\end{equation}
{the stationary condition is}
\begin{equation}
\frac{d}{dR}\left[\frac{R^2}{F(2R)}\right]=0.
\end{equation}
{Carrying out the differentiation gives}
\begin{equation}
2F(2R)=R F'(2R).
\end{equation}
{Substituting}
\begin{equation}
F(2R)=2\pi R a-\frac{4\pi\kappa}{R}-4\pi\gamma R,
\end{equation}
{and}
\begin{equation}
F'(2R)=2\pi a+\frac{4\pi\kappa}{R^2}-4\pi\gamma,
\end{equation}
{we obtain}
\begin{equation}
R_{\rm opt}=\sqrt{\frac{6\kappa}{a-2\gamma}}.
\label{eq:Ropt}
\end{equation}
{The condition for a real stationary radius is \(a>2\gamma\). A finite positive minimum wrapping time also requires \(0<g(R_{\rm opt})<1\). This condition is checked in the numerical examples.}

{The optimal radius decreases monotonically with increasing adhesion energy density. Stronger binding favors smaller particles because the increased driving force can overcome the membrane bending cost for smaller radii.}

{For physiologically relevant values of \(a\), Table \ref{tab:opt_size} lists the corresponding stationary radii. These values are stationary radii of the reduced time expression. They are optimal radii only when \(0<g(R_{\rm opt})<1\).}

\begin{figure}[H]
\includegraphics[width=0.5\textwidth]{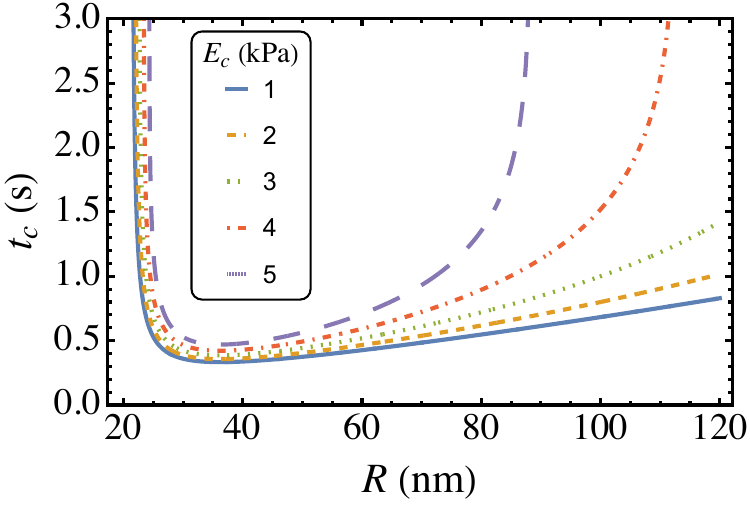}
\caption{Complete engulfment time \(t_c\) as a function of particle radius \(R\) for different host cell Young's moduli \(E_c=1,2,3,4,5\) kPa with \(E_v=100\) MPa. The stationary radius is independent of \(E_c\) in the reduced time expression. The finite-time domain depends on \(E_c\) and \(a\).}
\label{fig:wrapping_time_vs_R}
\end{figure}

\textls[-35]{For the HIV-1 gp120-CD4 parameter set used in the original manuscript, \mbox{\(a\sim 3.1\times10^{-4}\) J\,m\(^{-2}\)} gives \(R_{\rm opt}\approx43\) nm. This value is close to the radius scale of HIV-1 \cite{Sun2006}. The comparison is suggestive but does not establish evolutionary optimization. Viral size is constrained by virion assembly, genome packaging, envelope architecture, transmission fitness, immune pressure, and host-cell tropism. The result is proposed as a testable hypothesis. If depletion-mediated initiation is a significant selective pressure, engineered nanoparticles with radii near \(R_{\rm opt}\) should exhibit enhanced cellular uptake compared with particles of similar composition but different sizes.}

\subsection{Kinetic Phase Diagrams}

\subsubsection{Wrapping Time and Optimal Radius}

The complete engulfment time \(t_c\) from Equation \eqref{eq:t_c} is plotted as a function of particle radius \(R\) for several values of the cell Young's modulus \(E_c\) in Figure \ref{fig:wrapping_time_vs_R}. The curves exhibit a minimum at a stationary radius \(R_{\rm opt}\) that is independent of \(E_c\) in the reduced expression. As the cell becomes stiffer, the engulfment time increases and the size window \([R_{\min},R_{\max}]\) narrows. This non-monotonic behavior results from the competition between membrane bending, which is dominant for small \(R\), and cytoskeletal resistance, which is dominant for large \(R\). The divergence near the lower bound \(R_{\min}\) indicates that particles below this size cannot complete the modeled passive wrapping process in finite time. This statement is limited to the present model and parameter set. It does not imply that such particles cannot enter cells by other pathways.

\subsubsection{Phase Diagram in the \texorpdfstring{\((R,a)\)}{Ra} Plane}

A global view of engulfment feasibility is provided by the phase diagram in the \((R,a)\) plane (Figure \ref{fig:R_vs_zeta}), obtained from the cubic Equation \eqref{eq:cubic} for a fixed cell stiffness \(E_c=5\)~kPa. The shaded region between the lower boundary \(R_{\min}(a)\) and the upper boundary \(R_{\max}(a)\) defines the set of particle radii and adhesion energy densities that permit complete engulfment in the reduced model. The two boundaries merge at a limiting value of \(a\). This merger characterizes closure of the admissible size interval. The locus of stationary radii \(R_{\rm opt}(a)\) within the allowed region decreases with increasing \(a\), consistent with Equation \eqref{eq:Ropt} and Table \ref{tab:opt_size}. The two positive roots merge at a limiting parameter value. This is an algebraic folding of the boundary equation. It is not, by itself, a dynamical saddle-node bifurcation of the state variable \(h\).

\subsection{Critical Cell Stiffness}

{From the upper bound condition Equation \eqref{eq:upper_bound}, substituting \(F(2R)\) gives}
\begin{equation}
2\pi R a-\frac{4\pi\kappa}{R}-4\pi\gamma R
<\frac{32\sqrt2}{9}E_v R^2.
\end{equation}
{For a given particle radius and adhesion energy density, there exists a maximum cell stiffness above which the reduced model does not give a finite positive wrapping time. The expression is}
\begin{equation}
E_c^{\max}
=
\frac{9E_v F(2R)}
{32\sqrt2 E_v R^2-9F(2R)}.
\label{eq:Ec_max}
\end{equation}
{This expression is valid when \(0<F(2R)<32\sqrt2 E_v R^2/9\). At \(E_c=E_c^{\max}\), \(g=0\) and \(t_c\) diverges. This maximum is a critical upper limit for finite-time wrapping in the reduced~model.}

\begin{figure}[H]
\includegraphics[width=0.5\textwidth]{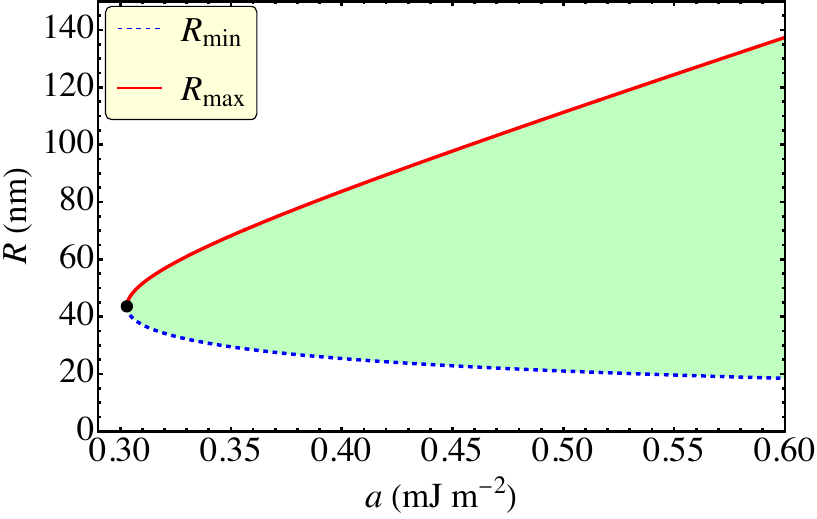}
\caption{Engulfment phase diagram in the plane of particle radius \(R\) and adhesion energy density \(a\) for \(E_c=5\) kPa and \(E_v=100\) MPa. The shaded region marks the complete engulfment window. The boundaries merge at a limiting parameter value. The stationary radius shifts to smaller values as \(a\)~increases.}
\label{fig:R_vs_zeta}
\end{figure}

\section{Discussion}

\subsection{Physical Interpretation and Variational Formulation}

The model provides a reduced two-stage description of receptor-mediated endocytosis. In the initiation stage, entropic depletion forces from molecular crowding facilitate the approach of the particle to the membrane. This provides one possible contribution to the initiation question that has been relatively overlooked \cite{Gao2005,Shi2008,Sun2006,Li2012}. In the engulfment stage, ligand--receptor binding sustains wrapping, and cytoskeleton viscoelasticity regulates kinetics through the Onsager variational principle. The competition among depletion attraction, binding, membrane deformation, and cytoskeletal resistance determines engulfment~feasibility.

{The correct planar depletion geometry within the Asakura-Oosawa model is used for the approach phase. It ensures volume continuity and recovers the flat-surface limit before contact. It also provides curvature corrections for the planar case. The extension of this planar volume into the wrapping phase is a separate approximation. The local parallel-plate adhesion density \(w_{\rm dep}=2Pr\) is used for the attached membrane in the reduced wrapping~model.}

{The variational formulation includes elastic storage and viscous dissipation of the cytoskeleton. The friction coefficient in the wrapping phase is \(\zeta_{KV}=8\eta_c\sqrt{Rh_H}/3\), which has the correct dimensions. The earlier expression proportional to \(h^{3/2}/\Phi\) had dimensions of force and was not a friction coefficient.}

The optimal size \(R_{\rm opt}=\sqrt{6\kappa/(a-2\gamma)}\) decreases with binding strength. Cell stiffness \(E_c\) modulates kinetics: higher \(E_c\) prolongs engulfment and shrinks the size window. These conclusions hold within the reduced model and the domain \(0<g(R)<1\).

\subsection{Comparison with Previous Models}

The model generalizes previous work in several limits. For \(c=0\) and finite \(E_c\), it reduces to elastic contact models of the type studied by Sun and Wirtz \cite{Sun2006} and Li \etal~\cite{Li2012}. For \(E_c\to0\) with fixed viscosity, the long-time elastic resistance of the cytoskeleton vanishes, and the thermodynamic membrane criterion of Yuan \etal~\cite{Yuan2010} is recovered. The optimal size relation generalizes receptor-diffusion models \cite{Gao2005, Shi2008} by explicit dependence on \(a\). The framework adds depletion forces and cytoskeleton viscoelasticity to membrane-shape studies such as Frey et al. \cite{frey2019dynamics}. The recent model of Shadmani et al. \cite{ShadmaniMehrafroozMontazeriRichards2026} shows that membrane tension can produce frustrated phagocytosis, a stalled engulfment state. The present model similarly predicts a finite engulfment window and a critical cell stiffness, beyond which engulfment cannot proceed in the reduced description. This consistency indicates a general role of membrane tension in limiting particle uptake.

{Other theoretical efforts have examined particle shape and local curvature in cellular wrapping \cite{KhosravanizadehSensMohammadRafiee2022, RichardsEndres2016}, as well as particle shape, orientation, and rotation during receptor-mediated endocytosis \cite{TangZhangYeZheng2018}. Experimental and simulation studies have demonstrated how particle shape dictates membrane interactions, from surfing to full engulfment \cite{VanDerHamAgudoCanalejoVutukuri2024}. These works complement the spherical-particle assumption used here and indicate directions for future generalization.}

\subsection{Biological Implications and Experimental Tests}

{The depletion mechanism operates without specific binding and provides one possible contribution to initial proximity. Depletion forces alone are relatively weak, approximately \(2\)--\(3\,k_BT\) at contact for the representative parameters. They are likely insufficient to overcome membrane bending resistance. Their primary role may be to reduce the effective barrier to contact and to increase the residence time of the particle near the membrane, thereby enhancing the probability of specific ligand--receptor binding.}

\textls[-15]{The creep retardation time \(\tau=\eta_c/E_c\) is \(1.348\) s for the representative parameters (see Table~\ref{tab:parameters}). It is long compared with fast endocytic events and may be relevant for slower processes. The predicted size window depends on \(a\), \(E_c\), \(E_v\), \(\kappa\), and \(\gamma\). The optimal radius for the HIV-1 gp120-CD4 parameter set is suggestive but not conclusive evidence of evolutionary~optimization.}

The phase diagram suggests experiments. Varying ligand density should shift the optimal size and broaden the window, consistent with Equation \eqref{eq:Ropt} and Table \ref{tab:opt_size}. Mechanical phenotypes of cells can probe the \(E_c\) dependence. {Direct force measurements during viral and nanoparticle uptake \cite{WiegandEtAl2020} provide an experimental avenue to test the predicted force-displacement relations.}

\subsection{Cluster-Mediated Uptake and Receptor Economy}

{Multiple particles can adhere into a cluster and be co-internalized. This possibility can be assessed within the present continuum framework using the minimum adhesion energy density \(a_{\min}(R)\) from Equation \eqref{eq:a_min}. For a cluster of \(N\) identical spherical particles of radius \(R\) that packs into a compact effective sphere, the effective radius is approximately \(R_{\rm eff}\sim N^{1/3}R\).}


For \(R=50\) nm particles with \(E_c=2\times10^3\) Pa, \(E_v=10^8\) Pa, \(\kappa=8.18\times10^{-20}\) J, and \(\gamma=2.07\times10^{-5}\) N/m \cite{Sun2006}, the individual-particle value is \(a_{\min}(R=50\,{\rm nm}) \approx 1.87\times10^{-4}\) J/m\(^2\). The function \(a_{\min}(R)\) has a minimum about \(1.829\times10^{-4}\) J/m\(^2\) near \(R\approx 59\) nm. For a dimer with \(N=2\) and \(R_{\rm eff}\approx 63\) nm, the required adhesion energy density decreases slightly, to about \(1.834\times10^{-4}\) J/m\(^2\). For an octamer with \(N=8\) and \(R_{\rm eff}\approx 100\) nm, however, the required adhesion energy density increases to about \(2.178\times10^{-4}\) J/m\(^2\). Thus clustering does not monotonically reduce the adhesion threshold. The engulfment window imposes an upper bound on the permissible cluster size. For the HIV-1 gp120-CD4 parameter set, the upper radius \(R_{\max}\) restricts \(N\) to a few. Cluster formation can therefore enhance receptor economy only within a limited size range set by the mechanical constraints of the engulfment process. The continuum model cannot address the extreme limit of a single receptor internalizing multiple particles. That limit requires a discrete stochastic treatment. The present estimate provides a testable quantitative prediction for experimental studies of virus cluster uptake.

\subsection{Limitations and Future Directions}

The present model has several limitations.

{First, the continuation of the planar depletion-volume formula into the wrapping phase is an approximation. The planar volume is exact only before contact. For \(h>0\), the membrane curvature changes the accessible volume for crowders. A fully self-consistent treatment would require solving for the membrane shape and computing the depletion volume for the curved geometry. This could be addressed using a Helfrich-based shape optimization coupled with the depletion interaction. The local parallel-plate adhesion density \(w_{\rm dep}=2Pr\) is used here as a reduced wrapping-phase approximation.}

{Second, the quasi-static viscoelastic approximation uses the step-response formula rather than the full hereditary integral. This is valid when the force varies slowly compared with the relaxation time \(\tau\). For fast engulfment events, the quasi-static approximation can misestimate the deformation. A full hereditary integral treatment would be more accurate and can be implemented with the auxiliary variable \(Q(t)\).}

{Third, the van't Hoff relation and the ideal hard-sphere Asakura--Oosawa model neglect non-ideal interactions, polydispersity, and spatial heterogeneity. The glycocalyx, membrane-associated macromolecules, and other crowders can modify the effective depletion interaction. For the representative hard-sphere volume fraction \(\phi\approx0.21\), non-ideal pressure corrections can be of order one. Replacing \(P\) by a non-ideal pressure alone does not provide an exact non-ideal depletion potential.}

{Fourth, the model treats the virus as approximately rigid and assumes uniform ligand and receptor distributions. Discrete bond stochasticity is neglected. Only spherical particles are considered. Multiple-particle adhesion and co-internalization are discussed only at a continuum level. Future work should consider multi-particle uptake and cluster formation.}

{Fifth, electrostatic interactions, van der Waals forces, and glycocalyx-mediated effects may also contribute to the initial approach. A more complete model would include these contributions. Recent work on how cells wrap around coronavirus-like particles using extracellular filamentous protein structures \cite{GuptaSantangeloPattesonSchwarz2025} highlights the role of additional adhesive structures beyond simple ligand--receptor binding.}

Future extensions should include particle shape effects \cite{KhosravanizadehSensMohammadRafiee2022, TangZhangYeZheng2018}, coupling of binding kinetics with deformation, membrane reservoirs, cytoskeletal network structure, and active fluctuations \cite{Song2025, Huang2013, Nividha2026enzyme}. Incorporating active signaling and actomyosin \mbox{contractility \cite{wu2018getting, wu2025generalized, Jacobson2019lateral}} would further enhance realism. {A promising direction is the explicit coupling of membrane tension with intracellular signalling cascades, as recently modeled in the context of phagocytic cup growth \cite{ShadmaniMehrafroozMontazeriRichards2026}. Such an extension would allow the model to capture tension-dependent biochemical regulation and its effect on engulfment kinetics.} {The theoretical frameworks developed for amoeboid swimming in confined geometries \cite{Wu2015, Wu2016, Farutin2019}, experimental studies on amoeboid swimming propelled by molecular paddling in lymphocytes~\cite{Aoun2020}, and chemotaxis-guided cell swimming \cite{Wang2021} can be adapted to model active cortical flows that accompany phagocytic cup formation and particle internalization.}

\section{Conclusions}\label{sec5}

{We have examined depletion attraction and cytoskeletal viscoelasticity in a reduced continuum model of particle-membrane association and wrapping. For ideal depletants, the sphere-wall excluded-volume construction gives an exact depletion potential for the planar geometry before contact. The potential and force vanish continuously at the onset of excluded-volume overlap. This interaction provides one possible contribution to membrane proximity before specific binding. Its extension to curved wrapping geometries requires additional approximation.}

{Within the adopted geometric assumptions and creep-based time approximation, the analysis yields a conditional minimum ligand density, criteria for a finite particle-size window, and stiffness-dependent limits to complete wrapping. When the stationary radius lies inside the domain of finite positive wrapping times, the estimated time has a minimum at \(R_{\rm opt}\), which decreases with increasing adhesion energy density. At fixed viscosity and other independent parameters, the same time approximation predicts slower wrapping as cell stiffness increases. The two positive roots defining the size window merge at a limiting parameter value, which characterizes closure of the admissible size interval.}

{Depletion attraction is interpreted as one possible contribution to particle-membrane association, alongside electrostatic interactions, steric effects, and membrane fluctuations. A concentration criterion based on comparison with thermal energy provides an energetic scale. The associated capture probability and residence time require a stochastic description. Quantitative predictions also require consistent treatment of curved-membrane exclusion geometry, recoverable elastic storage, and viscoelastic memory. The present analysis identifies how nonspecific attraction, specific adhesion, and mechanical resistance can contribute to different stages of membrane wrapping.}

\begin{acknowledgments}
Z.O. is supported by the Major Program of National Natural Science Foundation of China under Grant No. 22193032. H.W. is supported by the General Program of NSFC under Grant No. 12374210, the open research fund of Songshan Lake Materials Laboratory No. 2023SLABFN20, and the startup fund No. WIUCASQD2022005 from the Wenzhou Institute, University of Chinese Academy of Sciences.
\end{acknowledgments}

\section*{Data Availability Statement}
The original contributions presented in this study are included in the
article. Further inquiries can be directed to the corresponding authors.

\section*{Conflicts of Interest}
The authors declare no conflicts of interest.

\bibliographystyle{apsrev4-2}

\end{document}